\documentclass[final,5p,twocolumn]{elsarticle}

\usepackage{amssymb}

\biboptions{sort&compress}

\begin{document}

\begin{frontmatter}

\title{Size of the emission source and collectivity in  ultra-relativistic p-Pb collisions}

\author[agh,ifj]{Piotr Bo\.zek}
\ead{Piotr.Bozek@ifj.edu.pl}
\author[ujk,ifj]{Wojciech Broniowski}

\ead{Wojciech.Broniowski@ifj.edu.pl}

\address[agh]{AGH University of Science and Technology, Faculty of Physics and Applied Computer Science, al. Mickiewicza 30, 30-059 Krakow, Poland}
\address[ifj]{The H. Niewodnicza\'nski Institute of Nuclear Physics PAN, 31-342 Krak\'ow, Poland}
\address[ujk]{Institute of Physics, Jan Kochanowski University, 25-406 Kielce, Poland}

\begin{abstract}
The interferometric radii in the system formed in ultra-relativistic proton-lead collisions 
are investigated in a framework based on event-by-event 3+1~dimensional viscous hydrodynamics. 
We argue that the most central p-Pb collisions undergoing collective expansion behave 
similarly to the peripheral nucleus-nucleus 
collisions. The interferometric observables can serve as signatures of 
the formation of an extended fireball. 
\end{abstract}

\begin{keyword}
relativistic proton-nucleus collisions \sep relativistic hydrodynamics \sep interferometry \sep collective flow \sep LHC
\end{keyword}

\end{frontmatter}

The collective nature of the evolution of ultra-relativistic nucleus-nucleus (A-A) collisions from 
Relativistic Heavy-Ion Collider (RHIC) to the Large Hadron Collider (LHC) 
energies has been well evidenced, in particular with such phenomena as the harmonic flow or the 
transverse-momentum dependence of the interferometric radii. The description of the intermediate evolution stage
with relativistic viscous 
hydrodynamics yields a successful quantitative prediction at the level of, say, 15\% for the most 
relevant observables and for a wide range of centralities $c$, including rather peripheral collisions 
up to $c \sim 70$\% \cite{Shen:2011zc,Teaney:2009qa,Schenke:2011zz,Luzum:2011mm,Ollitrault:2012cm,newreview,Song:2008si}. The application of this collective approach to the proton-nucleus, not to mention the 
proton-proton (p-p) collisions, is more questionable \cite{Huovinen:2008te} 
but very intriguing \cite{Li:2012hc}, as some features typical for 
collective phenomena have been observed in the highest-multiplicity proton-lead (p-Pb) and p-p collisions as well.

\begin{figure*}
\begin{center}
\includegraphics[angle=0,width=0.6 \textwidth]{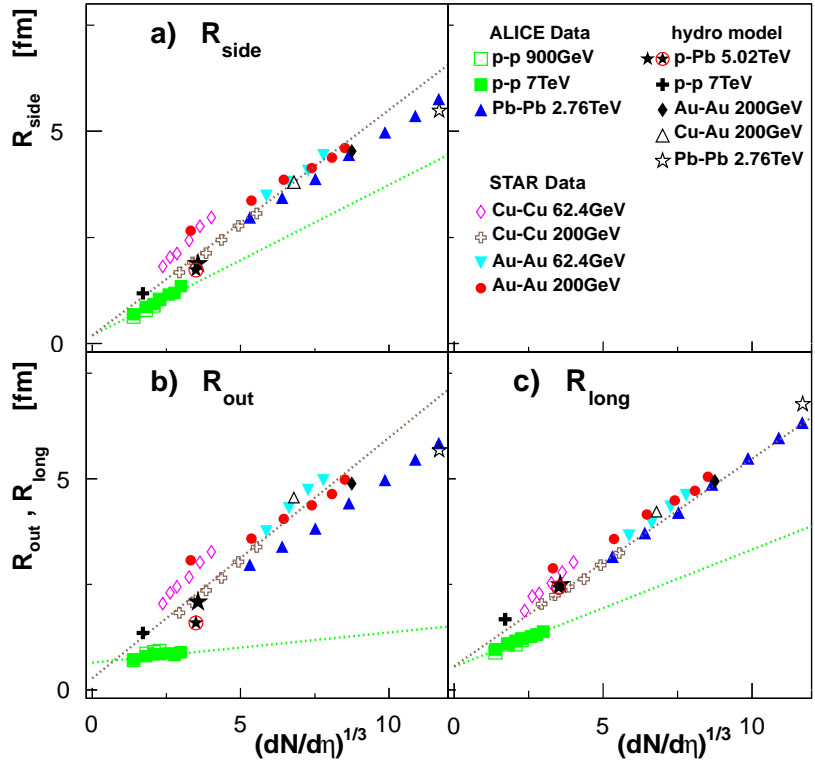}
 \end{center}
\vspace{-7mm}
\caption{Pion interferometry radii $R_{\rm side}$ (a), $R_{\rm out}$ (b), and $R_{\rm long}$ (c)
for average pair momentum $k_T=400$~MeV for different collisions systems and energies, 
plotted as functions of the charged particle multiplicity. The compilation of the experimental data 
is taken from~\cite{Kisiel:2011jg,Kisiel:2011jt}. 
The data for the RHIC energies are from the STAR Collaboration~\cite{Abelev:2009tp,Adams:2004yc} and 
for the LHC energies from the ALICE Collaboration~\cite{Aamodt:2011kd,Aamodt:2011mr}. The lines are to guide the eye. 
Various hydrodynamic calculations come from \cite{Bozek:2009dt,Bozek:2012hy,Bozek:2011ua}. 
The results of this work for the p-Pb system are indicated with filled stars (standard source) and encircled stars (compact source). 
\label{fig:hbtsys}} 
\end{figure*}  

The p-Pb collisions at $\sqrt{s_{NN}}=5.02$~TeV have been investigated at the 
LHC~\cite{ALICE:2012xs,ALICE:2012mj,CMS:2012qk,Abelev:2012cya,ATLAS:pPb},
with the original motivation to study the initial state effects
 ~\cite{Salgado:2011wc}
and, in particular, the saturation in QCD. 
Important observables in that respect are the two-particle 
correlations in relative pseudorapidity
and azimuthal angle. The long-range correlations in rapidity are formed in the very early 
stage of the collision and present
a dedicated probe of the initial state, on the other hand, the azimuthal correlations can be significantly modified  
by the final state rescattering. When a fireball of strongly interacting 
matter is formed, azimuthal correlations due to collective flow appear in the interaction 
region~\cite{Ollitrault:1992,Voloshin:2011mx}. For instance, in Ref.~\cite{Bozek:2012gr} we have 
argued that the appearance of 
the same-side ridge structure in the correlation 
data in the p-Pb collisions measured at 
the LHC is semi-quantitatively described with event-by-event hydrodynamics, which provides a strong case for the 
interpretation based on collective harmonic flow.  An alternative, 
appealing explanation of the appearance of the same-side ridge
in p-Pb interactions is based on the color-glass condensate approach for the initial state
\cite{Dusling:2012cg,Dusling:2012wy}. Therefore,
 it is important to look for independent estimates of the role of final state
 interactions in the dynamics.

In this Letter we argue that the behavior of the pionic interferometric radii in the most central p-Pb
 collisions should be used as a fingerprint
of collectivity: if the experimental results follow the pattern of A-A collisions, 
then collective behavior is 
present. This concerns both the values of the radii as well as their 
dependence on the transverse momentum of the pair, governed by the flow \cite{Akkelin:1995gh}. 
We evaluate the pionic interferometric radii for the most central p-Pb system with relativistic hydrodynamics and  
predict that in this treatment the system should closely imitate
the peripheral nucleus-nucleus collision. If this is confirmed experimentally, it should serve as another strong case
for the presence of collectivity in the most central p-Pb collisions.   
At the same time, the low-multiplicity p-Pb collisions are in our view not expected to display the above advocated
behavior and should follow the p-p pattern, hinting different dynamics. Typically, the size and life-time 
of the collective source formed in central p-Pb collisions is $3$-$4$~fm~\cite{Bozek:2011if}.

The initial geometry of a small-source system 
formed in p-Pb collisions is dominated by fluctuations, therefore the costly machinery of event-by-event 
viscous hydrodynamics must necessarily be applied to properly describe the azimuthally asymmetric components of the collective flow \cite{Schenke:2010rr}.
As the result, the hydrodynamic expansion in most central p-Pb 
collisions generates a sizable elliptic and triangular flow~\cite{Bozek:2011if}, while the two-dimensional 
correlation functions in relative pseudorapidity and azimuth  are 
in semi-quantitative agreement with 
experimental observations~\cite{CMS:2012qk,Abelev:2012cya,ATLAS:pPb}, following the  same
mechanism as in A-A collisions~\cite{Takahashi:2009na,Luzum:2010sp}.

Interferometric correlations
explored in this work are a snapshot of the emission points at the final stage~\cite{Lisa:2005dd,Wiedemann:1999qn}. 
Fitting the Bertsch-Pratt~\cite{Bertsch:1989vn,Pratt:1986cc} formula 
to the same-sign two-pion correlation functions yields estimates for the femtoscopic size of the emission source.
Quite remarkably, the systematics of the interferometry radii
in nuclear collisions at different energies shows an approximate extended scaling with
the multiplicity of the system~\cite{Aamodt:2011mr,Kisiel:2011jt},
\begin{equation}
R\propto \left(\frac{dN}{d\eta}\right)^{1/3} \, \, , \label{eq:scale}
\end{equation}
as shown in Fig.~\ref{fig:hbtsys}.
We note that for the A-A collisions the hydrodynamic results shown in Fig.~\ref{fig:hbtsys} are 
consistent with the data,  properly reproducing the scaling (\ref{eq:scale}) found experimentally. This 
stems from a generic relation between  the interferometry radii 
and the size of fireball in A-A collisions and
is found in many hydrodynamic calculations \cite{Broniowski:2008vp,Pratt:2008qv,Karpenko:2009wf,Soltz:2012rk}. 
Admittedly, more 
precise probes, sensitive to the details of the collective flow profile, 
such as the ratio $R_{\rm side}/R_{\rm out}$, make the exact agreement  
more difficult to achieve in hydrodynamics, however, with a proper choice of the initial condition and equation of state~\cite{Broniowski:2008vp,Pratt:2008qv,Karpenko:2009wf} it can
also be accomplished. The cascade model approaches also work properly for 
the A-A system~\cite{Lin:2002gc,Li:2012ta}.

\begin{figure}
\begin{center}
\includegraphics[angle=0,width=0.35 \textwidth]{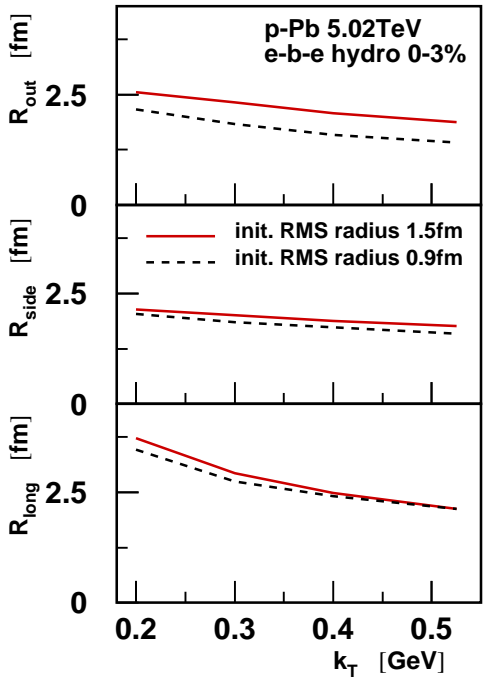} 
\end{center}
\vspace{-3mm}
\caption{The model predictions for the
pionic interferometric radii in p-Pb collisions for centrality 0-3.4\%, plotted as functions of the average
transverse momentum of the pair  for the standard source (solid lines) and compact source (dashed lines). 
\label{fig:hbtkt}} 
\end{figure}

On the other hand, the interferometry radii measured in p-p collisions show a similar scaling
trend but with a very different slope than the A-A case (cf.~Fig.~\ref{fig:hbtsys}). This indicates that the mechanism responsible for
the formation of the interferometric correlations may be distinct in  elementary \cite{Csorgo:2008ah} and  nuclear collisions.
We note that the hydrodynamic modeling of the p-p system assuming collective expansion and rescattering in the final state 
is not always compatible with the data~\cite{Bozek:2009dt,Werner:2010ny,Kisiel:2010xy,Li:2012np,Werner:2011yh}. 
We should admit that the analysis meets some difficulties: the extracted interferometric radii in the small p-p system 
depend strongly on the form of the fitted correlation function and the background subtractions, they require 
preservation of the conservation laws, 
moreover, depend on the resonance contributions or the effects of the uncertainty principle 
\cite{Aamodt:2011kd,Werner:2010ny,Kisiel:2010xy,Akkelin:2011zz}. Another important ingredient in small systems
that may influence  the interferometry radii and their momentum dependence is the unknown pre-thermal flow 
\cite{Karpenko:2009wf,Kisiel:2010xy}.
All in all, the hydrodynamic predictions for
the p-p pionic interferometric radii can overshoot the data by more than 50\%, showing that 
the hydrodynamic description
does not describe the p-p data in a uniform way.

The situation is hopefully different for the p-Pb system, where observed 
two-particle correlations \cite{CMS:2012qk,Abelev:2012cya,ATLAS:pPb} suggest 
the possible existence of collective flow. If this picture is true, it means
that the  system formed in p-Pb collisions is sufficiently large and long-lived
to accommodate a hydrodynamic expansion stage. 
In our model, the initial condition is generated with GLISSANDO~\cite{Broniowski:2007nz}.
The parameters of the calculations are similar as in~\cite{Bozek:2011if}, except that they are adjusted for 
the collisions energy of $\sqrt{s_{NN}}=5.02$~TeV. Thus we take the NN inelastic cross section 
$\sigma_{NN}=67.7$~mb, moreover, we use a realistic (Gaussian) wounding profile~\cite{Rybczynski:2011wv} for the NN collisions. 
When the individual NN collision occurs, a {\em source} is produced, meaning deposition 
of energy and entropy in a location in the transverse plane and spatial rapidity. 
In the conventional wounded nucleon model it is assumed that the sources are located in the transverse plane in the centers 
of the participating nucleons. This amounts to rather large initial transverse sizes in the p-Pb system. 
Locating the source in the center-of-mass of the NN system%
\footnote{We thank Larry McLerran and Adam Bzdak for this suggestion.} is also admissible, which leads to a more compact initial 
distribution. We use both variants, labeled {\em standard} and {\em compact}, which allows us to 
provide upper and lower limits for the size of the initial source and thus estimate the model uncertainty.
The average rms radii for the two sets of initial conditions are $1.5$ and $0.9$~fm, respectively.
The probabilistic nature of the Glauber model leads to initial source distributions which 
fluctuate event-by-event, 
referred to as the geometric fluctuations. No other possible sources of fluctuations in 
the initial phase, such 
as fluctuations of the color fields at smaller scales, are incorporated.

\begin{figure}
\includegraphics[angle=0,width=0.4 \textwidth]{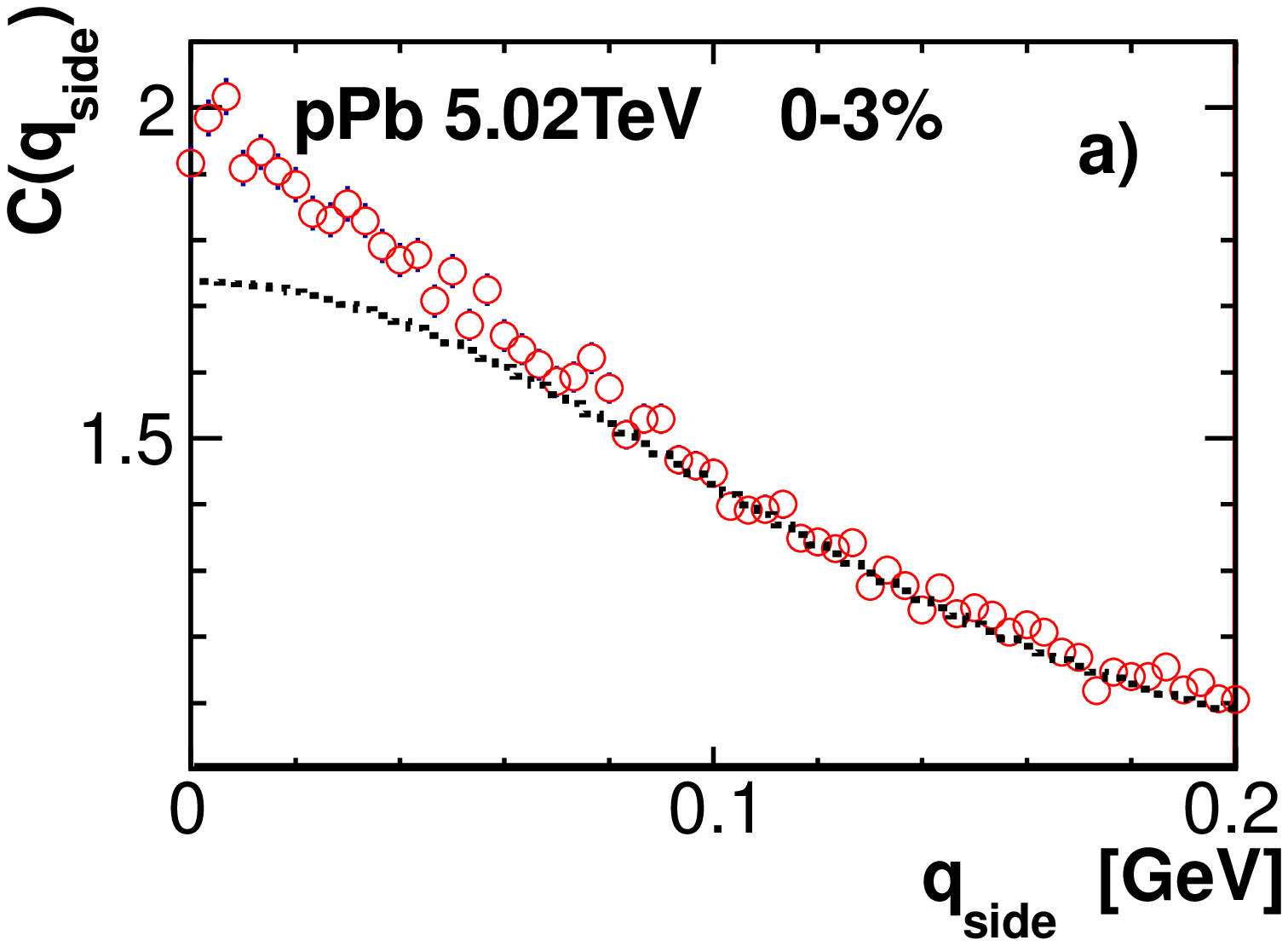} \vspace{-8mm} \\
\includegraphics[angle=0,width=0.4 \textwidth]{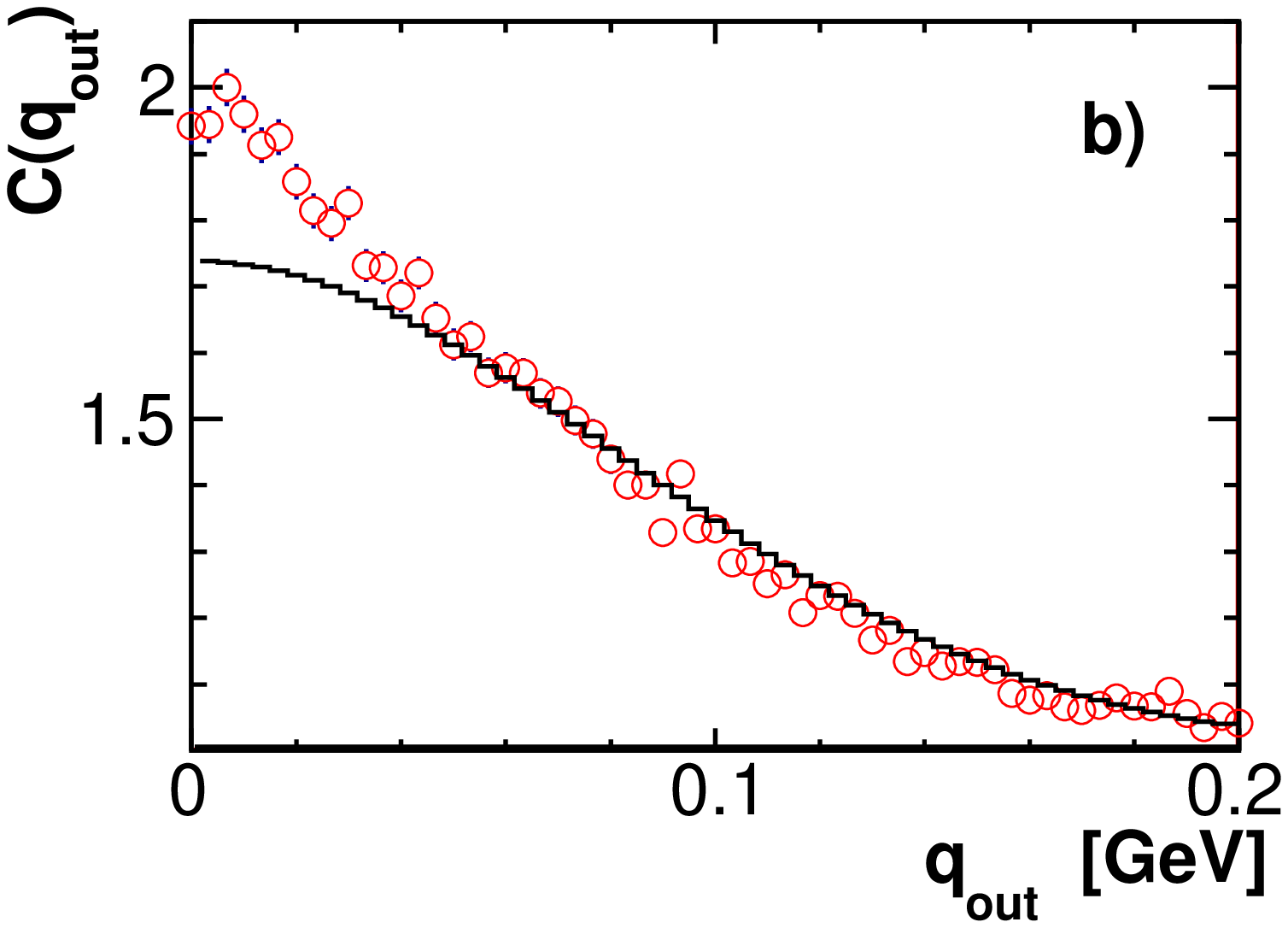} \vspace{-8mm} \\
\includegraphics[angle=0,width=0.4 \textwidth]{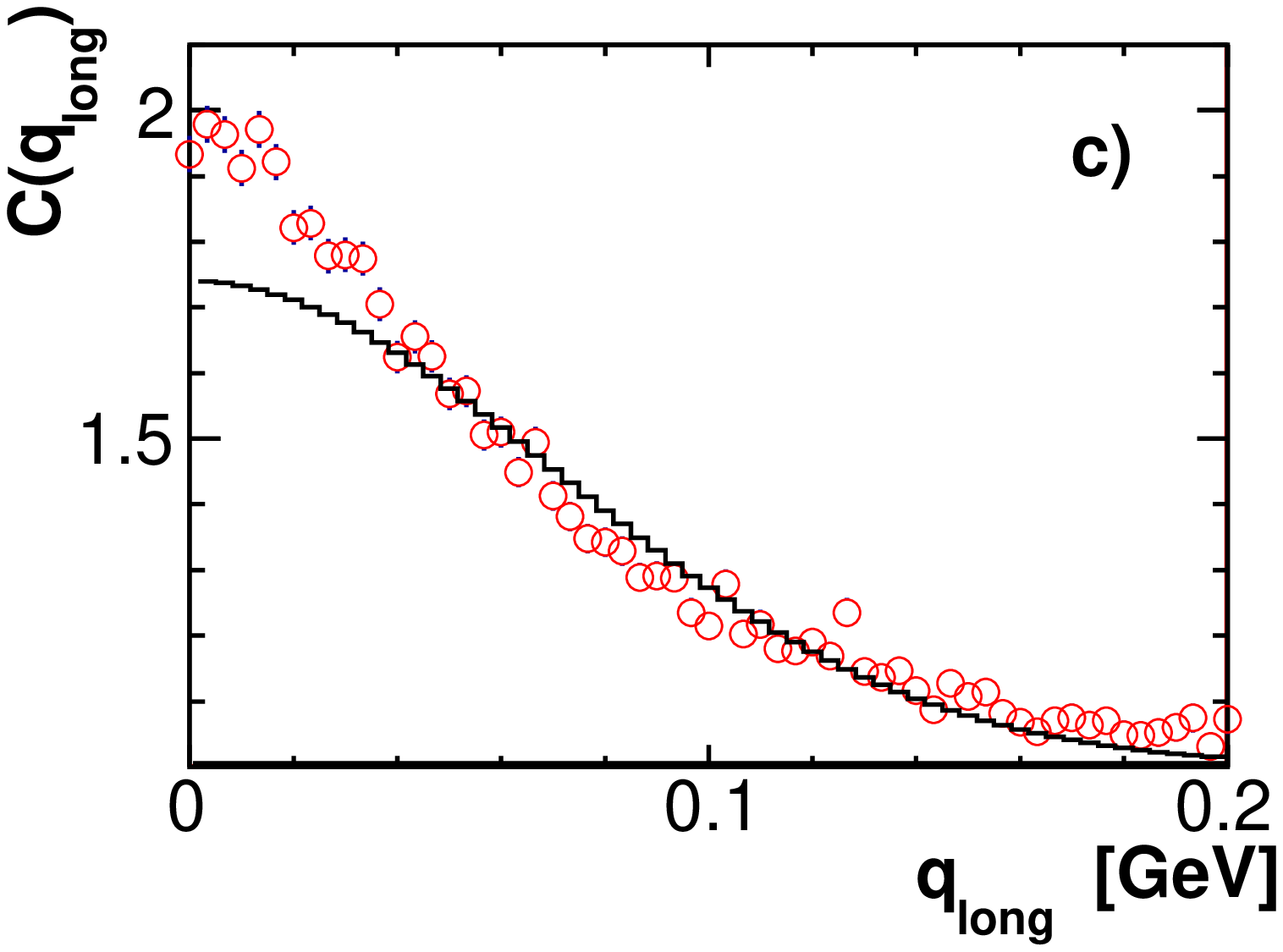} 
\vspace{-7mm}
\caption{The model predictions for the central sections, in direction $q_{\rm side}$, $q_{\rm out}$ and $q_{\rm long}$ 
(panels a), b) and c) respectively), of the interferometric 
correlation functions for the pion pairs (points) and the corresponding three-dimensional Gaussian fits
(lines) for the most central ($c=0-3.4$\%) p-Pb collisions. \label{fig:cor}}
\end{figure}

In our simulations we use the event-by-event $3+1$-dimensional viscous 
hydrodynamic model as described in detail in~\cite{Bozek:2011if,Bozek:2012gr}. 
%
The  statistical
emission and resonance decays at freeze-out are carried out with THERMINATOR2~\cite{Chojnacki:2011hb}. 
The centrality $0$-$3.4\%$ for the highest-multiplicity p-Pb collisions is 
defined in a simplified way by the condition on the number of participant nucleons, $N_{part}\ge 18$. We generate
$450$ distinct hydrodynamic evolutions of different initial conditions, both for the 
standard and compact source scenario. For each freeze-out 
hypersurface at $T_f=150$~MeV we generate 1000 
THERMINATOR events to increase the statistics.
The expansion of the compact source is faster than for the standard source. The stronger flow in the
compact source case causes an increase of the average transverse momentum by $20$\%, but the 
size of the fireball
at freeze-out is similar for the two considered scenarios.

The femtoscopic correlation functions for the pion pair are obtained with the two-particle method 
described in detail in Ref.~\cite{Kisiel:2006is} and implemented in THERMINATOR2~\cite{Chojnacki:2011hb}.
Technically, the correlation functions involve pairs from the same hydro events in the numerator,
and mixed pairs from different hydro events in the denominator of the correlation function
\begin{eqnarray}
\hspace{-7mm} C(q,k)=\frac{\int d^4x_1 d^4x_2 S(x_1,p_1)S(x_2,p_2)|\Psi(k^\ast,r^\ast)|^2}
{\int d^4x_1 S(x_1,p_1)\int d^4x_2 S(x_2,p_2)}\, , 
\end{eqnarray}
where  $q=p_1-p_2$ is the relative momentum of the pions, $k=(p_1+p_2)/2$ is the average pair momentum, and the asterisk indicates 
the variables boosted to the pair rest frame~\cite{Kisiel:2006is}.
For all the  THERMINATOR events generated from the same hydrodynamic event the emission source $S(x,p)$
is the same, 
hence pairs from such 1000 events can be combined in the model calculation to improve statistics in the numerator.  
Still, by combining pairs from different freeze-out hypersurfaces, or using the event-by-event averaged initial conditions, we find  that the 
effects of the event-by-event fluctuations 
of the emission source are small \cite{Bozek:2012hy}. 
The interferometric radii are obtained by fitting the Gaussian 
shape to the correlation functions. The Coulomb effects, expected to be very small in the p-Pb system, 
are not taken into account in the pair wave function $\Psi$, 
and consistently, no Coulomb corrections are used in the fitting procedure.

The result displayed in Fig.~\ref{fig:hbtsys} shows that our hydro predictions for the 
most central p-Pb system fall close to the A-A line, thus displaying the collective behavior. 
The distinction from the p-p trend is clearly seen, in particular for the standard-source case for $R_{\rm out}$ and  
$R_{\rm long}$, where the difference is about a factor of 2.
The compact source leads to somewhat smaller interferometric radii, in particular for $R_{\rm out}$ which is reduced by 
25\%. The other radii are very little affected by the initial source size.

In Fig.~\ref{fig:hbtkt} we give the dependence of the pionic femtoscopic radii 
on the transverse momentum of the pair, $k_T$. Again, we show the case of the most central p-Pb 
collisions, as these are most likely to display collectivity. We note the expected fall-off of the radii with $k_T$ 
caused by the collective flow.  For the compact initial distribution the values of the radii are above 1.8~fm for $R_{\rm side}$ and above 
2~fm for $R_{\rm out}$ and $R_{\rm long}$. 
For the compact case the $R_{\rm out}$ radius is visibly reduced, while $R_{\rm out}$ and $R_{\rm long}$ are only slightly smaller. 
Observing experimentally such large sizes 
(compared to the proton radius) in the most central p-Pb would 
demonstrate the formation of a fireball.

In femtoscopic studies the shape of the correlation function and its departure from the Gaussian form 
is frequently studied \cite{Kisiel:2010xy}, as it carries relevant information on the dynamics of the system. 
We provide this information for the standard source in Fig.~\ref{fig:cor}, where we plot the central sections of the 
correlation function. The departure from 
the three-dimensional Gaussian fit is clearly visible. 
In particular, 
at low momenta $q_a$ the correlation function is much sharper, rising well 
above the fitted Gaussian profile. Nevertheless, 
following the  experimental works reported in Fig. \ref{fig:hbtsys} we use the 
three-dimensional Gaussian profile in extracting the interferometry radii.

To conclude, we state again the importance of the experimental femtoscopic measurements for the p-A system, which 
will help to determine the nature of its dynamics. The proximity to the A-A scaling line of Fig.~\ref{fig:hbtsys} will place the system in the 
collective evolution mode, on the other hand, if it turns out to be close to the p-p line, elementary dynamics 
will be vivid. 
Our simple hydrodynamic calculation for the most central p-Pb system gives radii consistent with the A-A scaling. 
We should note, however, that 
in a more realistic treatment we expect some deviations due to remnants from the elementary p-p collisions, 
as modeled for instance in the 
core-corona picture. We also note that if a large size fireball is found, it could be used in quenching models 
to be compared with the $R_{AA}$ data.

\bigskip

Supported in part by Polish Ministry of Science and Higher Education, grant N~N202~263438, by 
National Science Centre, grant DEC-2011/01/D/ST2/00772, and  by PL-Grid Infrastructure.

\bibliography{../hydr}

\end{document}